\documentclass[
  journal=largetwo,
  manuscript=article-type,
  year=2020,
  volume=37,
]{cup-journal}

\usepackage{amsmath,amssymb}
\usepackage[nopatch]{microtype}
\usepackage{booktabs,journal_abbr_mac}

\definecolor{ref@color}{RGB}{0,76,153}
\usepackage{hyperref}
\hypersetup{colorlinks=true,linkcolor=ref@color,citecolor=ref@color,urlcolor=ref@color}

\title{Revisiting the bimodality of galactic habitability in IllustrisTNG}

\author{Ana Mitra{\v s}inovi{\' c}}
\affiliation{Astronomical Observatory, Volgina 7, 11060 Belgrade, Serbia}
\email[AM, BV]{amitrasinovic@aob.rs, bvukotic@aob.rs}

\author{Branislav Vukoti{\' c}}
\affiliation{Astronomical Observatory, Volgina 7, 11060 Belgrade, Serbia}

\author{Teodora {\v Z}i{\v z}ak}
\affiliation{Astronomical Observatory, Volgina 7, 11060 Belgrade, Serbia}
\alsoaffiliation{Faculty of Mathematics, University of Belgrade, Studentski trg 16, 11158 Belgrade, Serbia}

\author{Miroslav Micic}
\affiliation{Astronomical Observatory, Volgina 7, 11060 Belgrade, Serbia}

\author{Milan M. {\'C}irkovi{\'c}}
\affiliation{Astronomical Observatory, Volgina 7, 11060 Belgrade, Serbia}


\begin{document}

\begin{abstract}
{The potential of a galaxy to host habitable planets is one of the most important questions in astrobiology. It is tightly connected to the evolution of galaxy-scale properties and the underlying cosmological processes. Using the improved cosmological simulation IllustrisTNG, we revisit the claim that a population of small, metal-rich, star-forming galaxies (“Cloudlet”), forms a local peak on the mass-metallicity relation, reflecting an enhanced galactic habitability potential. We refine the earlier analysis by applying updated filtering criteria to identify a more refined sample, further selecting objects based on their history. This process resulted in a confirmed sample of 97 dwarf galaxies, alongside 519 additional structures of uncertain origin, potentially comprising both numerical artefacts and unrecognised physical systems. Under these stricter conditions, the proposed bimodality in galactic habitability is strongly diminished. However, the astrobiological potential of metal-rich dwarfs, most of which are compact remnants of more massive galaxies that underwent tidal stripping, is a thrilling area of exploration. Although dense stellar environments are traditionally seen as inhospitable, recent studies highlight the role of dynamic environments in enhancing the distribution of biological material. Furthermore, the potential habitability of tidal structures formed in the aftermath of galactic interactions is a fascinating possibility. Our findings suggest that non-traditional structures support conditions favourable for life, opening up exciting new avenues for astrobiological research. This research underscores the need for a holistic approach to studying habitability that moves beyond planetary and stellar-focused frameworks to incorporate the broader galactic environment. Understanding the interactions between galaxies, their evolution, and the influence of their surroundings is essential to developing a more comprehensive model of how and where life might emerge and persist across the Universe.}
\end{abstract}

\section{Introduction}

Habitability is the foundational concept of modern astrobiology, with the main research focus on the discovery and habitable conditions of Earth-like planets around nearby stars. As a natural extension of such research, there is a growing interest in habitability studies of galaxies as the building blocks of the Universe. Although the idea that different spatial locations within a galaxy might be more habitable than others is already prominent in the literature, the seminal work of \citet{Gonzalez+2001} was the first to fully conceptualise the Galactic Habitable Zone (GHZ), initially only for our own Galaxy. The concept was further refined shortly afterward by \citet{Lineweaver+2004}, and many more studies have followed \citep[e.g.][]{Prantzos2008, Gowanlock+2011, Spitoni+2014, 2016A&A...592A..96G, Vukotic+2016, Forgan+2017, Spitoni+2017, Gowanlock+gordon_habitability_2018, Kokaia+Davies2019, Cai+2021Galax, Spinelli+2021, 2023PASA...40...54M}. The vast amount of available observational data has allowed for the precise refinement of habitability-related constraints, with these studies focusing on the Milky Way. Some authors \citep{Carigi+2013, Spitoni+2014} also considered the habitability of our closest massive neighbour, the Andromeda galaxy, M31.

Early studies of exoplanet samples found that the giant exoplanet abundance is higher around stars with higher metallicity \citep[e.g.][]{2001A&A...373.1019S, Fischer+Valenti2005ApJ...622.1102F, 2005A&A...437.1127S}. Subsequent studies of more numerous samples of exoplanets argued that even the smaller planets are likely more common around metal-rich hosts \citep{2015ApJ...808..187B, Wang+Fischer2015AJ....149...14W, 2019ApJ...873....8Z,2020AJ....160..253L}.  

In the study of \citet{Suthar+McKay2012}, the authors have taken a more holistic and general approach to assessing the habitability of galaxies. They used a metallicity-based methodology as a precursor for the formation of planets with habitable potential \citep{Lineweaver2001} and applied it to two elliptical galaxies (specifically, M87 and M32). They found that both of them meet the criteria for habitability and argued that many galaxies, regardless of their morphological type, should support wide habitable zones. A study investigating the likelihood of different morphological types of galaxies hosting complex life \citep{Dayal+2015ApJ...810L...2D} determined that giant elliptical galaxies are the most probable habitable hosts. Similarly, according to the findings of \citet{Zackrisson+2016}, spheroid-dominated galaxies are more likely to host terrestrial planets. However, for a dissenting view about large early-type galaxies, see \citet{Whitmire2020}.

 At present, numerical simulations have become an indispensable tool for modern extragalactic astronomy and cosmology. These simulations aid in comprehending the evolution of galaxies and their interactions within galaxy clusters, among other scientific advances. Naturally, they have already made their way into the galactic habitability studies \citep[e.g.][]{Vukotic+2016, Forgan+2017, Vukotic2017n, Stanway+2018, Stojkovic+2019MNRAS, Stojkovic+2019SerAJ}. Understandably, cosmological simulations offer a unique window into how different factors and effects co-evolve on galaxy scales, helping to inform target selection strategies for future exoplanet and technosignature missions. As the search for life progresses from individual planetary systems to statistical astrobiology across galactic environments, it becomes increasingly important to understand which types of galaxies are likely to host habitable planets.

Cosmological simulations, in particular, and their merger trees were first applied in astrobiology by \citet{Stanway+2018} and were further explored by \citet{Stojkovic+2019SerAJ, Stojkovic+2019MNRAS}. More specifically, \citet{Stojkovic+2019MNRAS} analysed the original Illustris cosmological simulation, with emphasis on the Fundamental plane \citep{Mannucci+2010}, that is, the relation between stellar mass $M_\star$, gas-phase metallicity, and star formation rate (SFR). One specific manifestation of the Fundamental Plane is the mass-metallicity relation observed in the Local Universe (MZR, first reported in the seminal work of \citealt{Lequeux+1979}; see also \citealt{Djorgovski+Davis1987}). The authors discovered a significant number of dwarf galaxies that were metal-rich and located above the main sequence of this relation. They referred to this population as "the Cloudlet," a term we will adopt for the present study. The previous study used the original Illustris simulation, the only one publicly available at the time, which satisfied the volume and resolution criteria. However, the original Illustris simulation suite had issues with galaxy populations and tensions with observational data \citep[for a summary, see][Section 6]{Nelson+2015A&C}, with one of the inconsistencies related to the crucial parameter of metallicity. The latter is also evident in this study.

Moreover, some lower-mass objects might have formed through baryonic processes in already formed galaxies rather than due to structure formation and collapse, or they might have formed relatively late in the proximity of a more massive host, making them satellites at formation time with little to no dark matter content. These objects typically appear as outliers in most galaxy scaling relations, and it is advised that they be considered with caution \citep{Nelson+2015A&C}. Since the Cloudlet embodies the population of outliers in the well-established scaling relation, we will revisit this result using an updated cosmological simulation suite. In the meantime, the collaboration behind the Illustris project performed the so-called next-generation suite of cosmological simulations, commonly known as IllustrisTNG\footnote{Publicly available at \url{https://www.tng-project.org/data/}.} \citep{TNGmethods2017,TNGmethods2018,Nelson+2019ComAC}. As the successor to the original Illustris simulation, it uses an updated model with refinements that alleviated previous deficiencies and tensions with empirical data. Additional physics is also included, significantly expanding the scope and possibilities of this simulation suite.

Thus, revisiting the bimodality of galactic habitability using IllustrisTNG can help us determine the validity and physical basis of the previous result. Rather than seeking to identify additional candidates, our goal is to assess whether the Cloudlet population proposed in the previous work withstands more rigorous analysis and whether it represents a physically significant class of habitable galaxy hosts. We will revisit the hypothesis of the bimodality of galactic habitability using a stricter, more physically motivated set of filters that account for both dynamical reliability and evolutionary history. By correcting for known numerical artefacts and pitfalls, as well as by applying astrophysically grounded criteria, our study aims to clarify whether such a population of galaxies genuinely contributes to a second mode of galactic habitability. While simulations such as IllustrisTNG have advanced our ability to model galaxy populations and their internal properties, they are subject to mass resolution limits, numerical artefacts, and sub-grid modelling assumptions \citep[e.g.][]{Vogelsberger+2020NatRP...2...42V, Crain+vaddeVoort2023ARA&A..61..473C}. A critical examination of habitability trends in such simulations must account for these caveats. Additionally, it will enable us to consider metal-rich dwarf galaxies in a more nuanced manner, leading to new, valuable insights into the habitability of galaxies and larger-scale structures in general.

This work is organised as follows. In Section~\ref{sec:simsample}, we provide an overview of the fundamental parameters of the chosen IllustrisTNG simulation box and describe the criteria applied to filter our data sample. We also compare our selected sample with the previous study \citep{Stojkovic+2019MNRAS} and highlight the differences between the two samples, focusing on some numerical parameters and technical issues arising from the differences between the two simulation suites. Section~\ref{sec:metalrich}, representing the central quantitative part of our work, is focused on separating a subset of metal-rich dwarf galaxies and the necessary analysis of their relevant properties. In Section~\ref{sec:habitability}, we qualitatively explore the implications of our results on habitability by comparing them with previous work. We emphasise the need for a broader examination of habitability on a larger scale. Finally, in Section~\ref{sec:summary}, we summarise our work and give concluding remarks.

\section{Simulation and Sample}\label{sec:simsample}

The IllustrisTNG cosmological hydrodynamical simulations of galaxy formation were performed using the \texttt{Arepo} code \citep{Springel2010AREPO} and \citet{PlanckColab+2016} cosmological parameters:
matter density $\Omega_\mathrm{m} = 0.3089$, baryon density $\Omega_\mathrm{b} = 0.0486$, dark energy density $\Omega_\Lambda = 0.6911$, Hubble constant $H_0 = 0.6774\; \mathrm{km}\; \mathrm{s}^{-1}\; \mathrm{Mpc}^{-1}$,
power spectrum normalisation $\sigma_8 = 0.8159$ and a primordial spectral index $n_\mathrm{s} = 0.9667$. The simulations were initialised at redshift $z=127$, and the results are stored in $100$ snapshots from $z=20$ to $z=0$. The simulation suite consists of three flagship runs: TNG50, TNG100, and TNG300, named roughly after the side length of the simulation box (that is, $50$, $100$, and $300$ Mpc, respectively). In addition to differences in cubic volume, the three boxes also vary in particle resolution, serving different research purposes. The IllustrisTNG team notes that TNG100 \citep{Marinacci+2018, Naiman+2018, Nelson+2018, Pillepich+2018, Springel+2018} provides a perfect balance of volume and resolution. The mass resolution for the TNG100 simulation box is $7.5 \times 10^{6}\;\mathrm{M}_\odot$ for dark matter and $1.4 \times 10^{6}\;\mathrm{M}_\odot$ for baryonic particles. The volume, mass resolution, and the number of detected subhaloes (i.e. galaxies) at redshift $z=0$ are almost indistinguishable from the original Illustris simulation used in \citet{Stojkovic+2019MNRAS}, which makes it the most suitable option for this study.

Using the same filtering criteria as in the previous study, we will secure a reliable comparison that will enable us to draw accurate conclusions. This approach is essential to ensure that our findings are robust. Therefore, our filtering criteria are as follows. We consider only subhaloes with total mass $M \leq 10^{14}\;\mathrm{M}_\odot$, stellar mass $M_\star \geq 10^{7}\;\mathrm{M}_\odot$ and non-zero values of star formation rate (SFR) and stellar metallicity. Our sample size is nearly half the sample size of the previous work, with 62262 subhaloes. The disparity is likely due to the refined recipes in the IllustrisTNG model for star formation, chemical enrichment, and feedback, which affected SFR and metallicity filtering criteria. For instance, the original Illustris simulation had an issue with inefficient quenching of galaxies, leading to a high cosmic star formation rate density at lower redshifts, $z \leq 1$ \citep{Genel+2014, Vogelsberger+2014}.

\begin{figure}[ht!]
\centering \includegraphics[width=.99\columnwidth, keepaspectratio]{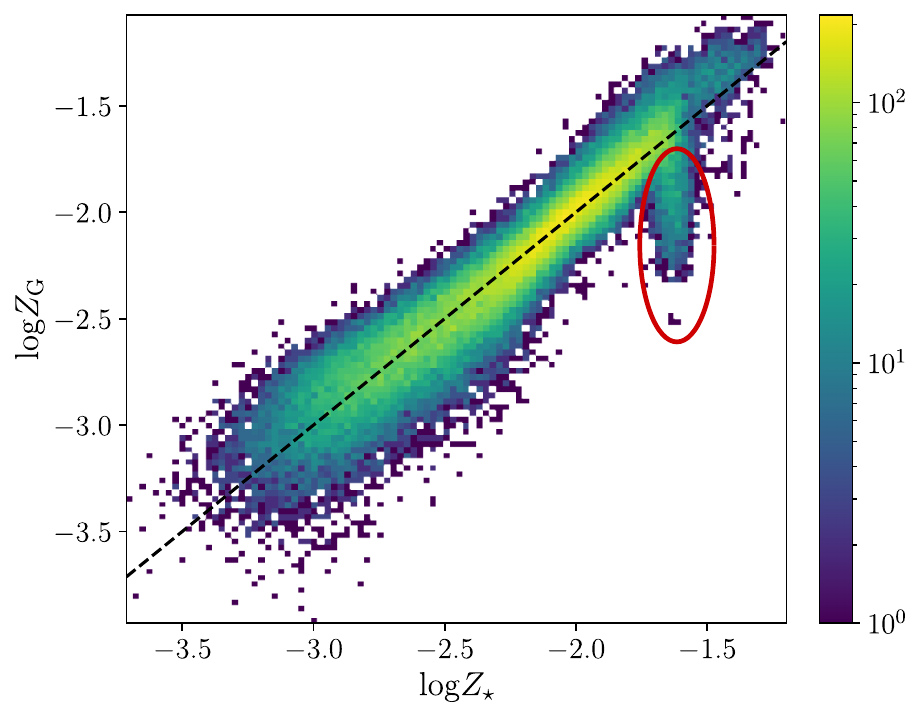}
\caption{Colour-coded number of subhaloes in the stellar metallicity - gas metallicity plane for the whole sample. The dashed black line represents the $y=x$ line, while the red ellipse highlights the non-negligible sub-population that is expected to reside below the observationally constrained MZR.}
\label{fig:ZsZg}
\end{figure}

This work will focus on stellar metallicity rather than gas-phase metallicity. The choice should not affect the results in any meaningful way, as there is an almost perfect linear correlation between the metallicities. Calculated for our sample, the Pearson correlation coefficient between these two variables is $\rho_\mathrm{p} \simeq 0.95$ with a zero $p$-value. Figure~\ref{fig:ZsZg} shows the visual representation of the correlation in the stellar metallicity - gas metallicity plane for the whole sample.

There is a distinct group of subhaloes with higher stellar metallicities and a range of lower, most likely sub-MZR, gas metallicities, which suggests a recent influx of metal-poor gas. A new study suggests that the inflow of metal-poor gas is primarily due to mergers \citep{perez-diaz_departure_2024}, although it can occur for various other reasons, depending on the environment. For astrobiological purposes and exploration of the present-day habitable potential, we are interested in the metallicity of an \emph{already formed} stellar population and its planetary systems. Hence, focusing on stellar metallicity is not only justified, but also more suitable for our goals. When exploring the relationship between stellar mass and stellar metallicity, we will refer to it as MZ$_\star$R to avoid confusion and misunderstanding with the observationally determined MZR based on gas-phase metallicity.

\subsection{SubhaloFlag Parameter}\label{sec:subhaloflag}

The \texttt{SUBFIND} algorithm \citep{SUBFIND2001} is widely used to identify gravitationally bound structures known as "subhaloes," which typically represent galaxies. However, as previously mentioned, not all these objects qualify as traditional galaxies formed due to structure formation and collapse. Special care is necessary for the original Illustris simulation, as it is impossible to differentiate between these genuine galaxies and potentially suspicious identified structures, mainly when dealing with lower-mass objects. Adding a new classifier\footnote{Detailed info is available at \url{https://www.tng-project.org/data/docs/background} in the "Numerical Considerations and Issues" subsection.} called \texttt{SubhaloFlag} has vastly improved the analysis process in IllustrisTNG. With this additional parameter, the team has simplified the filtering of objects to be more reliable.

The classifier can have two values: $0$ (\texttt{False}) and $1$ (\texttt{True}). The subhalo is flagged as \texttt{False} (or of non-cosmological nature) if all of the following criteria are satisfied: 
\begin{itemize}
    \item At its formation time, the subhalo is a satellite.
    \item The subhalo forms within the virial radius of its host.
    \item The dark matter fraction of the subhalo at formation is less than $0.8$.
\end{itemize}
The Illustris team emphasises that these are relatively conservative criteria adopted to ensure a low false-positive rate. As a result, some spurious subhaloes may not be flagged as \texttt{False} (these would be false negatives), and a more rigorous approach may require checking the history of the subhalo. It is essential to highlight that flagged subhaloes should not necessarily be treated as numerical artefacts. Strictly speaking, numerical artefacts stem from issues related to the code and other numerical recipes and solutions. Although some flagged subhaloes can fall under this category, others may represent genuine structures, although not galaxies (e.g. tidal debris or kinematically distinct substructures of larger galaxies). Another physically grounded possibility is that some of these flagged subhaloes represent genuine but exotic types of dark matter-deficient galaxies formed in the aftermath of a high-velocity head-on collision between two gas-rich galaxies \citep{Silk2019MNRAS.488L..24S, Shin+2020ApJ...899...25S, Lee+2021ApJ...917L..15L, vanDokkum+2022Natur, Lee+2024ApJ...966...72L}. In our qualitative analysis, we must acknowledge all possibilities when interpreting the results and avoid the hasty dismissal of flagged subhaloes (Section~\ref{sec:habitability}).

\begin{figure}[ht!]
\centering \includegraphics[width=\columnwidth, keepaspectratio]{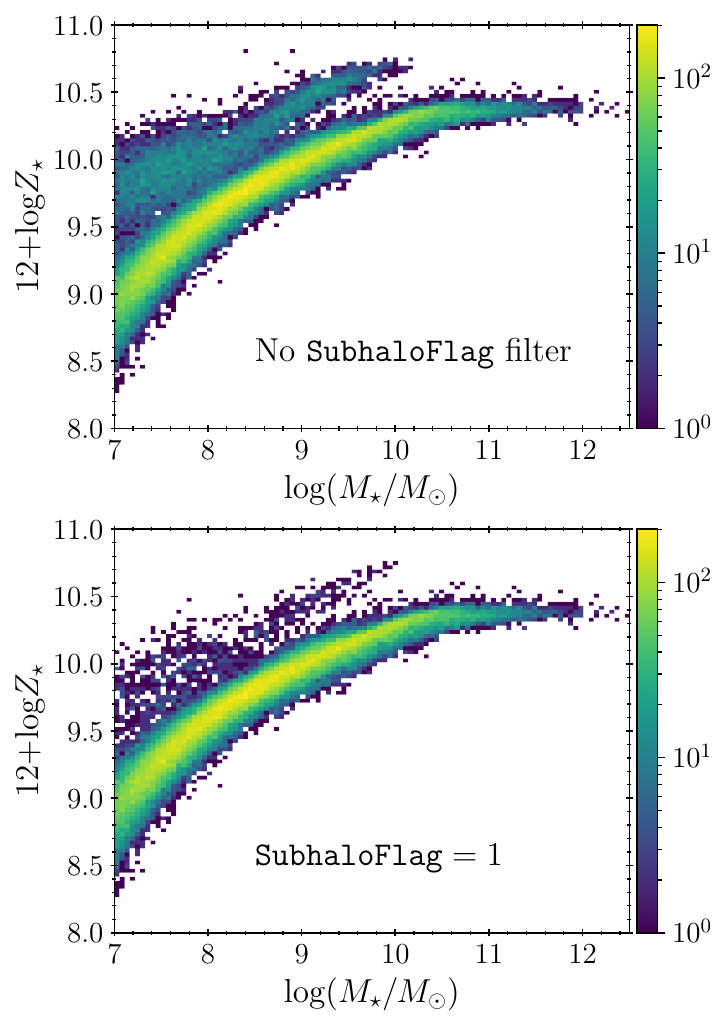}
\caption{Colour-coded number of subhaloes in the stellar mass-metallicity plane for the whole sample (upper panel) and the filtered sample for $\texttt{SubhaloFlag}=1$ (lower panel).}
\label{fig2}
\end{figure}

Since our sample includes subhaloes with both values of \texttt{SubhaloFlag} parameter, we show the color-coded number of subhaloes in the stellar mass-(stellar) metallicity plane in Figure~\ref{fig2} for the whole sample (upper panel) and additionally filtered \texttt{True} subhaloes (lower panel). The Cloudlet population is discernible in the upper panel, albeit appearing differently than in the previous work (i.e. there is a distinct second prong above the main sequence, but there is no noticeable, more significant gap between the two at lower masses). The additional filter clearly shows that this population predominantly comprises \texttt{False} subhaloes. However, even if we consider only genuine galaxies (lower panel), some scatter above the main sequence is still noticeable, although less significant.

As one of the primary aims of the present work is to revisit the previous result and reassess the habitability of genuine metal-rich dwarf galaxies in more detail, we will focus on this sub-population of the Cloudlet in the following. Naturally, we will also qualitatively discuss the possible habitability-related implications of the whole Cloudlet, including \texttt{False} subhaloes. A more detailed assessment of the habitability of the entire Cloudlet population requires a robust and precise classification of \texttt{False} subhaloes, which is beyond the scope of this work.

\subsection{Habitability: Bimodal Distribution of Galaxies?}
Previous work by \citet{Stojkovic+2019MNRAS} has found and named the Cloudlet population of metal-rich low-mass galaxies, implicating that it might be significant for its possible habitability aspect. Until then, studies of galactic habitability were indicating that large elliptical (and spiral) galaxies are most suitable for hosting habitable planets \citep[e.g.][]{Dayal+2015ApJ...810L...2D, Zackrisson+2016}, and the Cloudlet population presented a new and interesting possibility for life at the other end of the morphological spectrum. The significance of this hypothetical population of metal-rich dwarfs stemmed from their low SFR coupled with high (above-MZR) metallicity. The authors used the indicator given by \citet{Dayal+2015ApJ...810L...2D}, the quantity $N_\mathrm{p}$, which is widely used to assess the number of potentially habitable planets in different types of galaxies. This quantity scales with metallicity, which in general amplifies the probability of a habitable planet or moon forming, and is inversely proportional to SFR. Therefore, at a given (high) metallicity, Cloudlet galaxies have the distinct advantage of a lower SFR than their high-mass counterparts. The low mass of the Cloudlet galaxies decreases their overall number of potentially habitable planets or moons. Still, the expected high frequency of this type of galaxy was thought to compensate for the small number of habitable planets or moons hosted by each individual system.

While this earlier work presented a valuable concept, we revisit and expand upon their analysis using the updated IllustrisTNG simulation, which, as previously mentioned, features improved hydrodynamics, feedback models, and sub-grid prescriptions. Moreover, it offers improved agreement with observational constraints across a wide mass range. However, previous work contains several methodological limitations that significantly affect the interpretation of their results. Although the \texttt{SubhaloFlag} classifier, described previously, was not available at the time and was only introduced in the newer IllustrisTNG simulations, Cloudlet galaxies could have been systematically assessed for their physical nature or environmental conditions, which was not included in the study. For example, tracking the history of Cloudlet galaxies could have revealed that the majority of those objects formed extremely late during the simulation, perhaps even in the last few snapshots, which would point to the conclusion that those objects are not necessarily galaxies in a traditional sense. Examining the dark matter content and distances to larger and more massive galaxies could have also proven useful. Such metrics could have revealed whether a Cloudlet galaxy might in fact be a satellite, a substructure, or even a transient tidal feature. In summary, while the identification of a bimodality in galactic habitability was an intriguing result, we argue that the conclusions of \citet{Stojkovic+2019MNRAS} cannot be validated and should have been subjected to more rigorous scrutiny. In particular, the lack of an environmental, structural, and evolutionary context for Cloudlet galaxies undermines the robustness of the inferred bimodality. Our current study revisits this issue using a higher-fidelity simulation and a more stringent classification of galaxies.

\section{Metal-Rich Dwarfs}\label{sec:metalrich}

As this subsection should represent the central quantitative part of this work, we will have to carefully deal with the sub-sample of metal-rich dwarf galaxies, their significance, and their basic relevant properties. First, we will address the significance question using the leave-one-out cross-validation scheme in kernel density estimation in Section~\ref{sec:kernel}, which will give us insight into whether this scattered sub-sample is statistically significant to be considered further. Then, in Section~\ref{sec:dbscan}, we will employ the density-based clustering algorithm to isolate this sub-sample and clean it by checking the history of each subhalo and filtering out late-forming ones. In Section~\ref{sec:basicprops}, we will present some of the basic properties of the clean sub-sample, including but not limited to -- their compactness and the global rate of close encounters, their mass history, and morphological classification (where possible).

\subsection{Kernel Density Smoothing}\label{sec:kernel}

We applied kernel density smoothing to investigate the statistical significance of the separation between the high metallicity population and the bulk of the sample. Using a simple product Gaussian 2D kernel, the optimal smoothing bandwidths are calculated from the maximum likelihood cross-validation \citep{Duin1976}. 

\begin{figure}[ht!]
\centering \includegraphics[width=0.98\columnwidth, keepaspectratio]{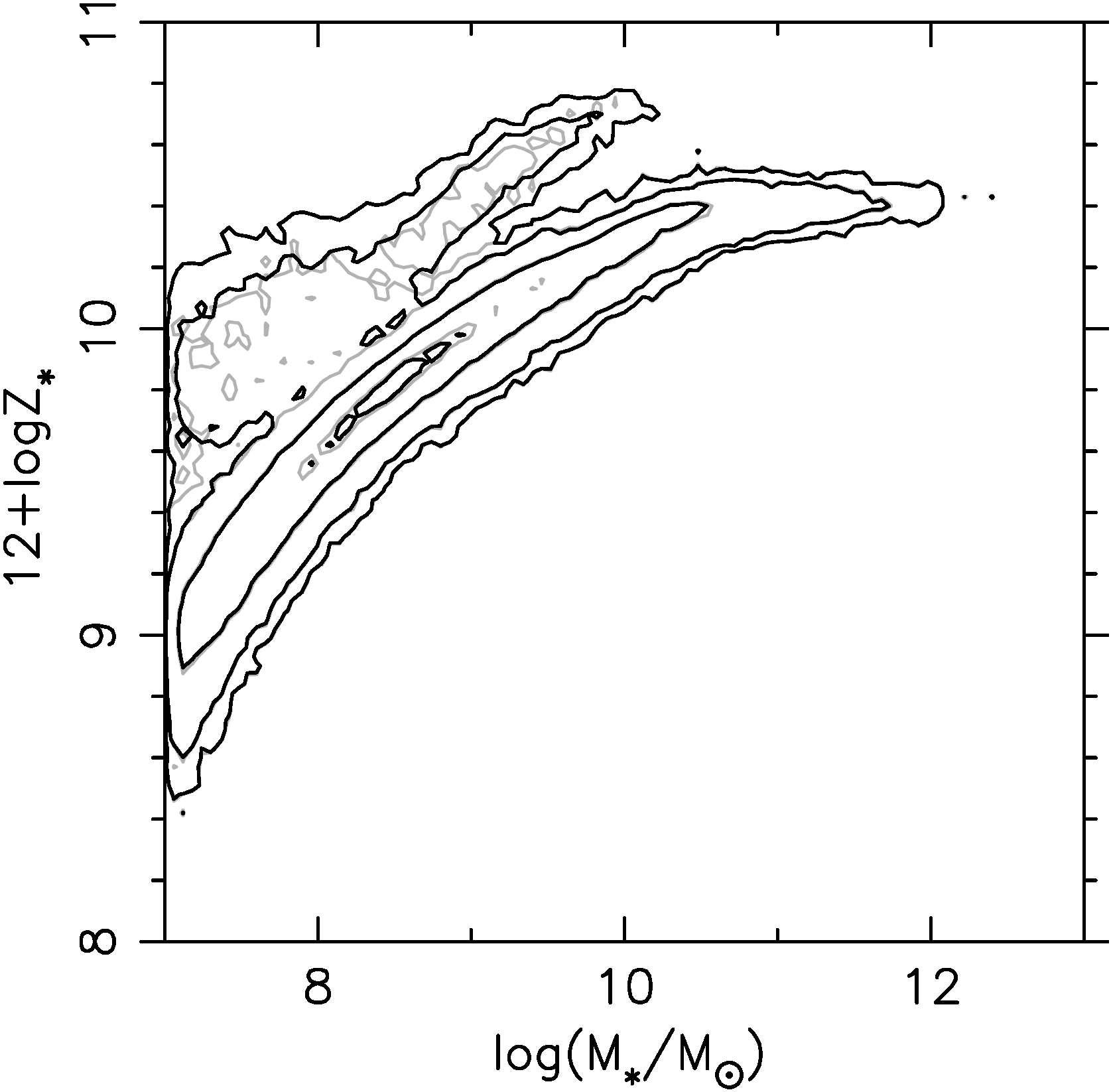}
\caption{Kernel density smoothing and the resulting probability density distribution for the selected sample in the metallicity-mass plane. Contour levels are at 0.01, 0.05, 0.5, and 1.5. Black -- all data points (kernel widths $h_{\log M}=0.029$, $h_{\log Z}=0.016$), grey (kernel widths $h_{\log M}=0.029$, $h_{\log Z}=0.017$) -- points with subhalo flag value 1.  }
\label{fig:kernel_density}
\end{figure}

Figure \ref{fig:kernel_density} shows the resulting probability density distributions. The total sample and sub-sample cases show a similar distribution in the bulk of the sample. At the same time, the high-metallicity segment is separated from the rest of the sample for $\log (M_\star/M_\odot)> 8.5$ and $12+\log Z_\star > 10$. The separation is visible at the first contour level of $0.01$ for the subhalo sub-sample, while the whole sample indicates the separation even at the $0.05$ contour level. The subhalo objects, therefore, have less pronounced separation compared to the rest of the data points from the sample. 

Given that the highest contour value is $1.5$, the separation occurs at best at the few per cent level for the whole sample and up to the $1\%$ level for the subhalo objects. In the $\log (M_\star/M_\odot)< 8.5$ and $12+\log Z_\star < 10$ region, there is no clear separation, and the high-metallicity population can rather be considered as the extended tail of the main population. Given that there are no other visible separations within the sample, it is evident that apart from the main branch of the sample, there is a significantly smaller branch, yet visible at a per cent level, in the $\log (M_\star/M_\odot)> 8.5$ and $12+\log Z_\star > 10$ region.

\subsection{Density-Based Clustering}\label{sec:dbscan}

\begin{figure}[ht!]
\centering \includegraphics[width=0.98\columnwidth, keepaspectratio]{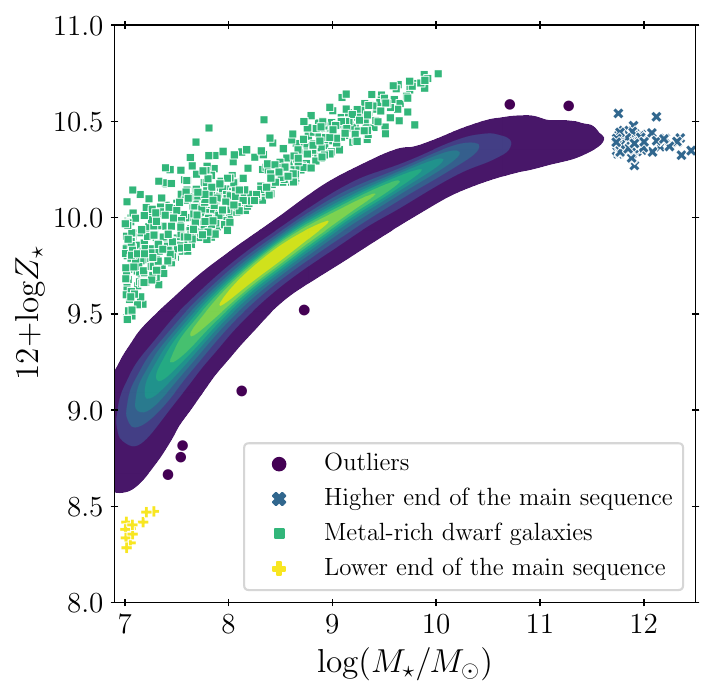}
\caption{Results of the \texttt{DBSCAN} algorithm application in the stellar mass-metallicity plane. Contours represent the main sequence, while the scatter points represent detected clusters, whose labels are indicated in the legend.}
\label{fig:dbscan}
\end{figure}

To identify metal-rich dwarf galaxies scattered above the main sequence of the mass-metallicity relation, we used a density-based clustering algorithm called \texttt{DBSCAN} \citep{Ester1996ADA, Schubert+2017DBSCAN} included in the \texttt{scikit-learn} package\footnote{For full documentation, see \url{https://scikit-learn.org/stable/modules/generated/sklearn.cluster.DBSCAN.html}.} \citep{scikit-learn}. The algorithm has two crucial input parameters: \texttt{eps} and \texttt{min\_samples}. The former defines the maximum distance between two samples that should be considered neighbouring points. The latter sets a threshold for the minimum number of samples in a neighbourhood. Although \texttt{eps} is considered the most critical parameter in determining the outcome for a given dataset, \texttt{min\_samples} controls the average density of the resulting detected clusters (the higher the value of \texttt{min\_samples}, the denser the detected clusters will be, and vice versa). This algorithm is most effective for datasets with distinct clusters of similar density, which is not the case with our sample. However, alternative algorithms that can detect clusters of various densities and sizes failed to detect the sub-sample of metal-rich dwarf galaxies as a single cluster, which was our desired outcome. We describe our procedure in the following and show the results in Figure~\ref{fig:dbscan}.

First, we apply the \texttt{DBSCAN} algorithm using $\texttt{eps}=0.1$ and $\texttt{min\_samples}=50$ as input parameters. This allows us to isolate the main sequence of the stellar mass--(stellar) metallicity relation, MZ$_\star$R. The results revealed a single cluster (covering almost the entire main sequence) and outliers. Objects in the sparse regions of the main sequence (i.e. its higher and lower ends) are recognised as outliers. We then apply the algorithm to a sub-sample of outliers, with $\texttt{eps}=0.2$ and $\texttt{min\_samples}=10$, which were adjusted to the new data. This revealed new data clusters: higher and lower ends of the main sequence, metal-rich dwarf galaxies scattered above, and true outliers. The results are shown in Figure~\ref{fig:dbscan}, and the sub-sample of metal-rich dwarf galaxies contains $616$ subhaloes, about $1\%$ of the whole sample.

To ensure that the sub-sample is composed of genuine metal-rich dwarf galaxies, we also trace the history of each subhalo and record the snapshot in which the subhalo was formed (i.e. when it first appeared). The results, with the subhalo counts per snapshot bin, are listed in Table~\ref{tab:firstsnap}. Interestingly, most of the subhaloes in this sub-sample appear for the first time in the last snapshot at redshift $z=0$. This makes them false negatives of the procedure that assigns the \texttt{SubhaloFlag} described in Section~\ref{sec:subhaloflag}. Only $97$ subhaloes have formed sufficiently early, at redshifts $z \geq 7.6$. We consider them genuine metal-rich dwarf galaxies and will explore their relevant basic properties in the following.

\begin{table}[ht!]
\caption{Snapshot number (i.e. \texttt{SnapNum}) and redshift $z$ at formation time of each subhalo in the sub-sample of metal-rich dwarfs. The total number of subhaloes (i.e. Count) is given per bin.}
\label{tab:firstsnap}
\begin{tabular}{|c|c|c|c|c|c|}
\toprule
\texttt{SnapNum} & $1-9$ & $66$ & $81-90$ & $91-98$ & $99$ \\
\midrule
Redshift $z$  & $14.99-7.6$ & $0.52$ & $0.24-0.11$ & $0.1-0.01$ & $0$ \\ 
\midrule
Count        & $97$ & $1$ & $9$ & $48$ & $461$ \\
\bottomrule
\end{tabular}
\end{table}

\subsection{Basic Properties of Metal-Rich Dwarfs}\label{sec:basicprops}

One of the crucial conditions for habitability is the continuity of habitable-friendly conditions, the so-called "habitable time" defined by \citet{Vukotic+2016}. Close stellar encounters can disrupt this continuity if they are catastrophic enough to alter the orbits of planets located in circumstellar habitable zones. The rate of close encounters $\Gamma$ is calculated as $\Gamma =  n_\star \langle v \rangle \langle \sigma \rangle$, where $n_\star$ is stellar number density, $\langle v \rangle$ relative stellar velocity, and $\langle \sigma \rangle$ cross-section of the encounter having adverse consequences. Although the formula is quite simple, its application to numerical simulations is not entirely straightforward. This is mainly due to the stellar number density, which is not readily available and requires assumptions about the average stellar mass \citep[as discussed by][]{2023PASA...40...54M}. We can thus roughly estimate the average global rate of close encounters for each metal-rich dwarf galaxy if we make the appropriate assumptions. Alternatively, we can first check how compact these galaxies are. One solution would be to use one of the well-known and observationally determined criteria \citep[e.g.][]{Barro+2013ApJ...765..104B, Damjanov+2015ApJ...806..158D}. However, these observationally-based criteria may be too restrictive, and an alternative approach would be more appropriate. This approach is based on the idea that compact galaxies are outliers on the mass-size relation \citep[used by, e.g.][]{Lohmann+2023MNRAS.524.5266L}, the relation which is well reproduced by IllustrisTNG \citep{Genel+2018MNRAS.474.3976G}. 

\begin{figure}[ht!]
\centering \includegraphics[width=0.99\columnwidth, keepaspectratio]{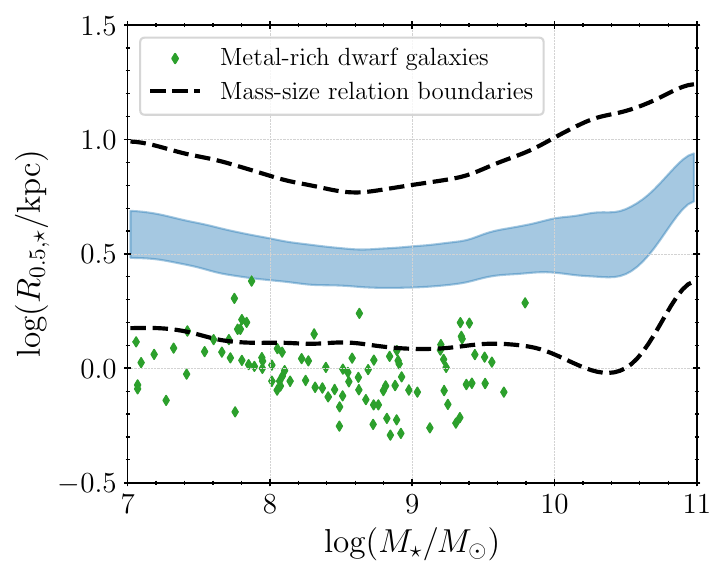}
\caption{Mass-size relation in the TNG100 simulation box, determined considering all the galaxies in the stellar mass range $7 \leq \log (M_\star/\mathrm{M}_\odot) \leq 11$. The expected boundaries of the relation are represented with black dashed lines, while early-formed metal-rich dwarf galaxies are marked with green diamonds.}
\label{fig:MSR}
\end{figure} 

The mass-size relation in the TNG100 simulation box, along with its boundaries and metal-rich dwarf galaxies, is shown in Figure~\ref{fig:MSR}. With a few exceptions, most of the early-formed metal-rich dwarf galaxies are indeed compact. Such systems are densely populated, which in turn should give a high rate of close encounters $\Gamma$. What is expected to affect and increase the rate of close encounters even further is the fact that these systems are metal-rich, making average stellar mass most likely sub-solar, which would result in higher stellar number density. Hence, it is likely that, in these galaxies, the rate of close encounters is too high.

\begin{figure}[ht!]
\centering \includegraphics[width=0.99\columnwidth, keepaspectratio]{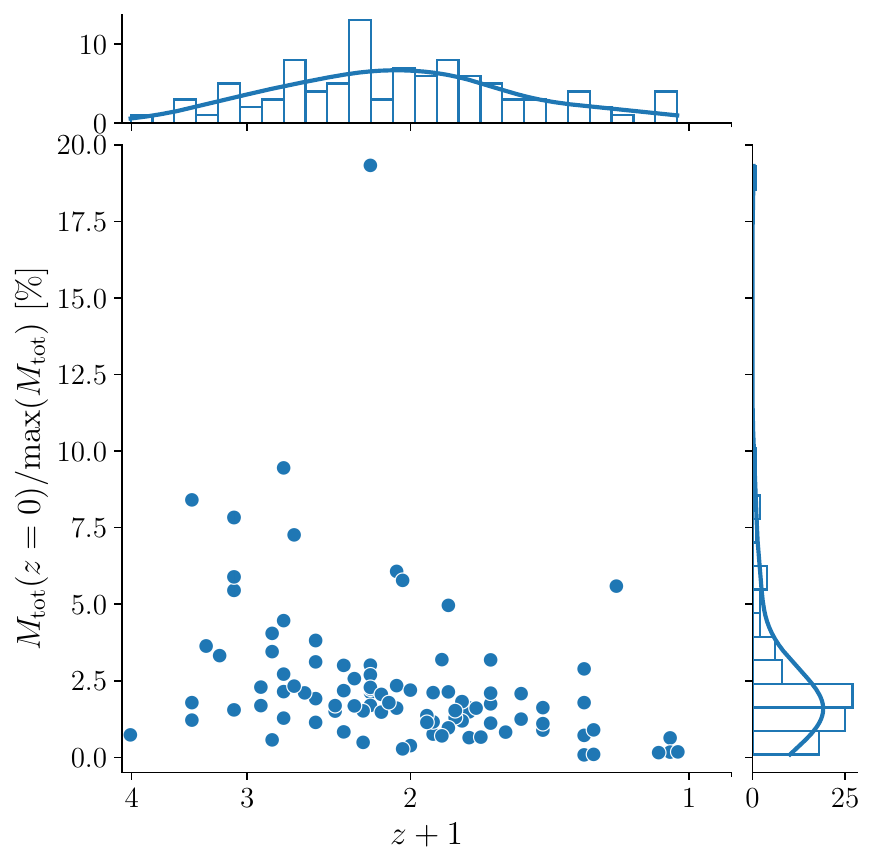}
\caption{History of the total mass of metal-rich dwarf galaxies, where the $x$-axis represents the redshift when the galaxy reached its maximum total mass, and the $y$-axis represents present-day total mass as a percentage of the maximum total mass. Histogram counts are given for each axis, with the corresponding probability density function.}
\label{fig:mass-history}
\end{figure}

Since most of the galaxies in our clean sub-sample of metal-rich dwarfs are compact, the next question is how these systems evolved to become compact. To answer this, we start by looking into the mass history of each galaxy in this clean sub-sample. In Figure~\ref{fig:mass-history}, we show present-day total mass (thus, including the gaseous and dark matter components) as a percentage of the maximum total mass and the redshift when the galaxy has reached its maximum mass. By redshift $z=0$, all of the galaxies in this sub-sample have retained very little of their maximum acquired mass. This suggests that the whole sub-sample is composed of compact, or almost compact, dwarf galaxies that have formed through the tidal stripping process, a highly plausible formation pathway, as demonstrated by numerous studies \citep[e.g.][]{bekki2001, bekki2003, pfeffer2013, pfeffer2014, martinovic2017, fm2018, kim2020, Deeley+2023MNRAS.525.1192D, Lohmann+2023MNRAS.524.5266L, Moura+2024MNRAS.528..353M}. The stripping formation pathway implies that the compact dwarf galaxies are tidally stripped remnants of much more massive galaxies processed by dense environments (e.g. galaxy clusters). Moreover, since the tidal stripping is an outside-in process \citep[e.g.][]{Diemand2007, Choi2009}, it primarily affects the extended dark matter halo component of galaxies, resulting in lower dark matter fractions in these galaxies in general. In the clean sub-sample, the median dark matter fraction is about $f_\mathrm{DM} \simeq 0.5$, and as many as $92$ out of $97$ galaxies have dark matter fractions lower than the cosmic average of about $f_\mathrm{DM} \simeq 0.84$ \citep{PlanckColab+2016}. Hence, the fact that the galaxies in the clean sub-sample are outliers on the mass-metallicity relation can be interpreted from an evolutionary point of view. The galaxies were, most likely, not always outliers but moved out of the main sequence as they lost a portion of their mass while retaining generally higher metallicities. According to \citet{Paul+2017MNRAS.471....2P}, clusters of galaxies are groups more massive than $8\times 10^{13} \; \mathrm{M_\odot}$. Following that definition and setting a lower mass threshold for a galaxy group of $5\times 10^{12} \; \mathrm{M_\odot}$, we find that only $4$ galaxies in our sample are located in clusters, $17$ are members of groups, while the rest (majority of the sample, $76$ galaxies) are satellites of more massive galaxies.

\begin{figure}[ht!]
\centering \includegraphics[width=\columnwidth, keepaspectratio]{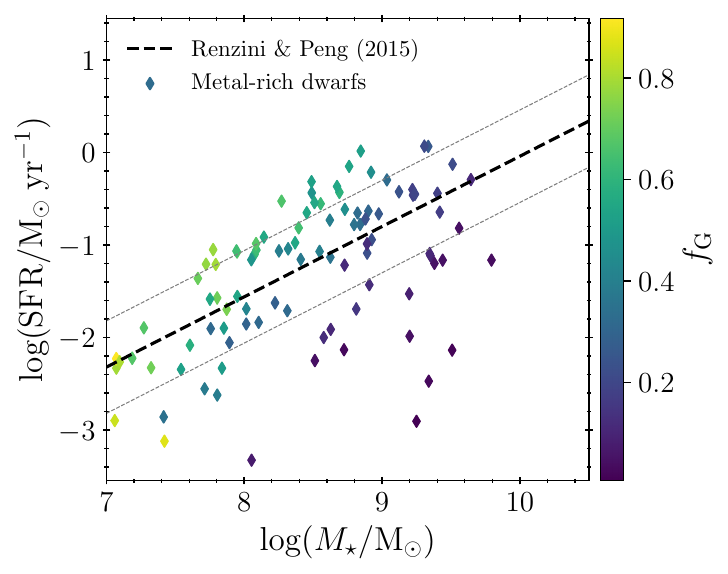}
\caption{Star-forming main sequence \citep[][black dashed line, where the dotted lines represent the boundaries $\Delta=\pm 0.5$]{Renzini+Peng2015ApJ...801L..29R}, and the metal-rich dwarf galaxies in the stellar mass -- star formation rate plane, colour-coded for the baryonic fraction of gas.}
\label{fig:sfrMS}
\end{figure}

The present-day stellar photometric data for the SDSS bands, particularly of interest $g$ and $r$, are available in the supplementary catalogue \citep{Nelson+2018}. The median colour of the clean sub-sample is quite blue $(g-r) \simeq 0.2$, while $78$ out of $97$ galaxies can be classified as blue with $(g-r)<0.4$, indicating a significant population of younger stars. Morphological information is also available in one of the supplementary catalogues \citep{Genel+2015ApJ...804L..40G}, allowing at least a rough classification corresponding to the Hubble sequence. However, for the TNG100 simulation box, the catalogue is restricted to galaxies that are sufficiently well resolved for a task, with stellar mass $M_\star > 3.4 \times 10^8 \; \mathrm{M}_\odot$. The information on morphological decomposition is, thus, only available for $45$ out of $97$ galaxies in the clean sub-sample. Most of them can be classified as elliptical as the median fraction of the spherical component is high, $f_\mathrm{spherical} \simeq 0.74$, while $6$ of them can be classified as irregular due to a relatively high fraction of the stellar mass that is not accounted for (i.e. it is neither part of the spherical components nor a thin disc). We emphasise that most galaxies are elliptical, not necessarily perfectly spherical, since their flatness, calculated based on the provided eigenvalues of the mass tensor, ranges between $\sim 0.68$ and $\sim 1$, where perfectly spherical galaxies have flatness equal to $1$. This implies that various sub-classes of elliptical galaxies are present in the sub-sample. One of the galaxies, which has a sufficiently high thin disc fraction, $f_\mathrm{thin} \simeq 0.19$, and the lowest flatness, can arguably be classified as lenticular, depending on the definition. Figure~\ref{fig:sfrMS} shows where the metal-rich dwarf galaxies are positioned in the stellar mass -- star formation rate plane, concerning the observationally-determined star-forming main sequence \citep{Renzini+Peng2015ApJ...801L..29R}. The galaxies are colour-coded based on their baryonic fraction of gas, most of which lie within the main sequence (within the boundaries of $\Delta=\pm 0.5$). A fraction of mostly gas-rich galaxies are barely above, while the quenched galaxies scattered below the main sequence are typically gas-poor.

\section{Habitability Implications}\label{sec:habitability}

The present analysis implies that a small population of low-mass galaxies is more metal-rich than those that form the main mass-metallicity correlation. The high-mass end of this population is separated from the main part of the sample, while the lower-mass end is connected to the main sample as if it is the extended tail of the main distribution. The members of this metal-rich Cloudlet with masses $\gtrsim10^9$\, M$_\odot$ can be considered as a separate branch and, as such, might provide habitability traits that differ from the main branch of galaxies, especially since metallicity is one of the main habitability agents and a necessary building block for Earth-like planets, which are also likely to exist around metal-rich stellar hosts.

However, whether or not this relatively small population of dwarf galaxies might be considered superior in habitability terms in comparison to the majority of galaxies is a complex question. In this section, we will discuss the galactic habitability assessment as it relates to this study, attempt to re-evaluate the habitability of metal-rich dwarfs, and qualitatively discuss the implications of this study in a broader perspective, taking into account the Cloudlet in its entirety.

\subsection{The Complexity of Galactic Habitability}

Galactic habitability is a multifaceted concept, shaped by both intrinsic properties of galaxies, essentially unknowable initial conditions at their formation, and their evolutionary history. Typically, the evolution of galaxies strongly depends on their immediate environment and external factors. Thus, galactic habitability depends on a number of properties that may bolster or hinder habitable conditions, and these effects are often strongly interrelated. 

External factors often influence crucial internal properties of galaxies. This is the most evident in dense environments, such as galaxy clusters, where close interactions and mergers are the most frequent, disrupting the stability of stellar systems, causing starbursts, multiple reignitions of the nuclear engine, and other perturbing phenomena. However, even in a quiescent environment, stellar orbits in a Milky Way-like spiral galaxies are subject to perturbations within the broad stellar neighbourhood; and even stars with nearly circular and stable orbits are partly coming from populations that have radially migrated from other parts of the galactic disk \citep{2023PASA...40...54M}. Tidal stripping and dynamical changes due to galactic flybys and interactions, which also occur in isolated environments, although less frequently, can produce a wealth of substructures \citep[e.g.][]{kim2014, Pettitt+Wadsley2018, Mitrasinovic+Micic2023}, making the habitability assessment much more complex than the original zonal habitability concept \citep{Gonzalez+2001}. Similarly, strong tidal stripping (associated with dense environments, interactions, or massive neighbouring galaxies) can indicate a dynamically unstable environment, further reducing the potential for sustained habitability. 

As mentioned previously, tidal stripping can lead to the formation of ultra-compact dwarf galaxies. In such densely packed stellar systems (i.e. compact dwarf galaxies in this case), close interactions between individual stars occur frequently, significantly impacting the potential habitability of planets. These interactions make it less likely for planets to maintain stable orbits within the habitable zones around their host stars. On the other hand, a high density of the stellar system offers potential for interplanetary transport of biological material \citep[lithopanspermia,][]{Wesson2010SSRv..156..239W}, as argued by \citet{Balbi+2020Life...10..132B} and interstellar panspermia in general \citep{vukotic2021planet}.

While proxies such as high stellar metallicity and low SFR have been historically used to infer favourable conditions for planet formation and long-term stability \citep[e.g.][]{Lineweaver2001, Lineweaver+2004, Dayal+2015ApJ...810L...2D}, such an interpretation is perhaps overly simplified. On the one hand, the high star formation rate could be considered a habitability risk factor: a higher SFR would imply a higher rate of supernova and gamma-ray burst explosions, which could be detrimental to the development and survival of life on a planet's surface. Additionally, a high SFR coupled with generally low gas content might imply that the gas reservoir in these galaxies will soon be depleted, resulting in a smaller number of newly formed habitable systems. However, the effects of a high SFR could be positive: high-energy radiation, stemming from supernova explosions, while potentially dangerous, is highly unlikely to completely wipe out all life on the surface of a planet \citep{Sloan+2017NatSR...7.5419S}; however, it could provide the necessary energy for complex prebiotic chemistry, or fuel photosynthesis \citep{Lingam+Loeb2019MNRAS.485.5924L}, or even trigger genetic mutations in surface life that would allow more complex life forms to arise. Moreover, studies of mass extinction events on Earth \citep[e.g.][]{McElwain+Punyasena2007TEcoE..22..548M, Alroy2008PNAS..10511536A, Filipovic+2013SerAJ.187...43F} have led to the conclusion that these catastrophic events can lead to even greater biodiversity and may be necessary for the development of complex life.

Ultimately, any claim about habitability advantages must account for this multi-dimensional complexity. However, we have to specify what we are considering when discussing habitability: the emergence of life and development of complex life forms, or their sustainability and persistence. At first glance, these concepts appear very similar, but they can be entirely different. Factors that are considered detrimental to life development and sustainability are sometimes the same ones that can lead to the formation of complex life forms and the transportation of biological material across planetary or stellar systems.

\subsection{Re-evaluating the Habitability of Metal-Rich Dwarfs}

While the original Cloudlet population appeared promising due to its high metallicity and relatively low SFR, our refined analysis reveals a more nuanced picture. In light of the discussion above, we revisit the potential habitability of galaxies identified in our final filtered sample. Genuine dwarf galaxies examined in this work are metal-rich, which could imply superior habitability conditions, and the effects of their star-formation rates suggest considerable habitable potential. However, these galaxies are overwhelmingly compact, exhibit signs of strong mass loss over time, and typically represent remnants of formerly more massive systems that experienced tidal stripping. In many cases, they lie near massive host galaxies, suggesting that environmental processes played a dominant role in shaping their present-day properties. Although such galaxies may, on average, show favourable values in habitability proxies such as high metallicity and low SFR, these alone do not guarantee suitability for life. The stellar encounter rate, which scales with stellar density and internal dynamics, is generally high for these compact systems. This points to potentially disruptive environments, where planetary orbits may be unstable on longer timescales, which may inhibit long-term habitability. The existence of high metallicity alone is not sufficient to conclude a habitable advantage, especially in dynamically violent environments.

As discussed previously, dynamically challenging environments are not necessarily inhospitable to life. Various factors that affect habitability are intertwined, but they can also have opposing effects individually. Further studies of metal-rich dwarf galaxies are required to closely examine these opposing effects and their joint effect on habitability. Unfortunately, this is beyond the scope and possibilities of the present work. Due to resolution limitations, the explored galaxies contain only a few hundred stellar particles at the very most, making robust orbital perturbation analysis unfeasible. Future work exploring this interesting avenue should be done using detailed analysis based on higher-resolution data or tailored zoom-in simulations.

For the sake of argument, we will assume that the sample of Cloudlet galaxies examined in this work is indeed superior in their habitable potential. Compared to the original Cloudlet population, which probably included many young or dynamically inconsistent objects, these galaxies represent a much smaller and more physically meaningful subset. Importantly, even if these systems offered favourable conditions, their relative abundance is extremely low. Only 97 early-forming, filtered metal-rich dwarf galaxies remain in our final sample out of tens of thousands of subhaloes. This suggests that, even under optimistic assumptions about planetary formation, the global contribution of such systems to cosmic habitability is negligible. 

These results stand in contrast to the conclusions of the previous study \citep{Stojkovic+2019MNRAS}, which suggested that our search for extraterrestrial life should not be limited to galaxies similar to the Milky Way, but should also include these high-metallicity dwarfs. Genuine dwarf galaxies, as they are conventionally understood and examined in the present study, constitute only a small fraction of the Cloudlet population. This could lead us to the conclusion that previous assumptions about these objects may have been overly optimistic. Instead, we find that the Cloudlet is not a distinct secondary peak in the galactic habitability landscape, but a sparse tail shaped by extreme evolutionary pathways. We therefore conclude that the bimodality of galactic habitability reported in earlier work does not persist when evaluated in a more physically consistent simulation framework. Although some systems may retain marginally habitable properties, they do not constitute a distinct or significant population.

\subsection{Perspective Beyond Conventional Hosts}

In this study, we focused on clearly defined and dynamically coherent galaxies. Our results suggest that the previously identified Cloudlet population, when carefully filtered for reliability and formation history, does not represent a robust or dominant component of the habitable galaxy population. However, this naturally raises the question of what kinds of structures were excluded by our conservative criteria. In particular, the unfiltered Cloudlet of metal-rich low-mass objects is predominantly composed of structures that are not considered galaxies in a traditional sense. It is quite possible that a significant portion of this unfiltered population, much larger in numbers than our analysed sample, represents distinct substructures of larger, more massive galaxies. As such, it provides an additional strong argument for the complex habitability assessment that differs from the pioneering annular zone concept \citep{Gonzalez+2001}. The complexity should be intuitively understandable, given that galaxies are not as centrally influenced (in terms of radiation and motion) as individual planetary systems and their host stars. More detailed models of galactic habitability, especially focusing on galaxy dynamics and taking into account different sources of high-energy radiation, coupled with exoplanet hunt observations in nearby Milky Way satellites and other members of our Local Group, are crucial to provide better insights in future research on this subject. The localised nature of habitable regions within galaxies is already accepted and commonly discussed qualitatively, although the traditional view of the GHZ as a simple annular region \citep{Gonzalez+2001, Lineweaver+2004} remains a norm in quantifiable terms (e.g. when the study aims to constrain habitable regions of a galaxy). Thus, it should not be controversial that distinct substructures of larger galaxies may show more habitable-friendly conditions. In fact, this could be an additional argument that challenges the traditional model, urging us to re-examine our approach to large-scale habitability.

At the same time, the remaining Cloudlet population is composed of diverse objects, not just substructures of larger, more massive galaxies. Although some of these objects may be numerical artefacts, it is highly plausible that most of them are either dense, weakly bound tidal debris or previously mentioned, kinematically distinct substructures of larger galaxies. Both possibilities offer exciting and fresh insights, providing a new perspective on the holistic approach to habitability. Regarding tidal structures, their habitable potential was already speculated, although not thoroughly investigated, by \citet{Forgan+2017}. It should be noted that all of the above-mentioned objects are dynamically and evolutionarily challenging, which could affect the habitable potential by influencing the long-term stability of orbits. However, as we discussed previously, it should not be considered detrimental to life as it might appear at first glance. There is clearly a balance between the long-term stability required for life to develop and thrive and the dynamic, maybe even violent, nature of an environment required for spreading biological material across the galaxy and over larger scales in general.

Finally, these structures merit further investigation, and a broader follow-up will be needed to explore whether these ambiguous objects and structures might nonetheless host conditions conducive to life. In follow-up work, we plan to analyse these ambiguous or transitional structures and objects in more detail, incorporating, perhaps, if possible, particle-level dynamics and environmental analysis. This will help us better understand their origin, nature, degree of perturbation, and survival. It might very well be that, instead of genuine metal-rich dwarf galaxies, these ambiguous or transient structures are the ones that the previous study identified as possible cradles of life. Even if they prove to be unstable environments for life to persist and develop, they could hold a key to spreading life across large scales.

\section{Summary and Conclusion}\label{sec:summary}

We have revisited the interpretation of metal-rich dwarf galaxies in cosmological simulations as potential cradles of life, focusing on the population previously identified as the Cloudlet in the original Illustris cosmological simulation. This study employs the use of the improved IllustrisTNG, which gave an updated sample based on refined filtering criteria; further filtering of our own, based on the history of each subhalo in the sample, left us with 97 genuine dwarf galaxies, vastly outnumbered by the remaining 519 structures whose physical origin remains uncertain. We expect a certain percentage of them to be numerical artefacts; however, some (perhaps even the majority) could be genuine physical structures, and their origin calls for further investigation.

The final sample suggests the existence of a separate population of metal-rich dwarf galaxies above the main sequence of the mass-metallicity plane; these galaxies appear to branch away as the mass increases. Most of them will likely have more massive progenitors stripped of their gas content in galactic interactions, resulting in a lower-than-average gas content with varying star formation. Dark matter fractions are much lower than the cosmic average and further support the tidal stripping scenario.

The habitability assessment of these objects is not straightforward. Higher star formation rates and dense stellar environments traditionally point to lower habitable potential; however, recent studies of habitability of densely populated regions, as well as research on the impact of metallicity on the formation of habitable planets, suggest an altogether different possibility. Rocky, Earth-like planets could be relatively common in these galaxies, and their dynamical environment may serve to their advantage when it comes to the transportation of biological material. Similar reasoning could be applied to tidal structures, objects that we did not examine in detail in the present study, that formed recently in the aftermath of highly dynamic and violent events (e.g. interactions or mergers of galaxies). This is not the first time tidal structures have emerged as possible habitats, so it might be the perfect time to finally entertain the idea that structures other than conventional galaxies may be suitable for life hosting and development.

Our results suggest that the metal-rich Cloudlet population identified in a prior study is not representative of a large, habitability-favourable galaxy class. The majority of genuinely compact, metal-rich dwarfs are remnants of tidal stripping and are dynamically challenging environments for life. Moreover, these galaxies are exceedingly rare when proper filtering is applied. The apparent bimodality in galactic habitability suggested by \citet{Stojkovic+2019MNRAS} does not hold up under closer scrutiny using improved simulations and methodology. Thus, we conclude that there is no robust evidence for a second peak in galactic habitability associated with metal-rich dwarf galaxies. Although such systems may still host
habitable planets in specific cases, they are not statistically significant compared to massive spirals or ellipticals in both their overall number and in the number of hosted habitable stellar systems. These findings refine our understanding of how habitability is distributed across galaxy types and highlight the importance of careful simulation-based filtering. Future work will examine the broader population of metal-rich but unbound or ambiguous substructures, which may provide further insight into non-traditional hosts of life.

Evidently, addressing the question of habitability in the Universe requires a shift beyond planetary or stellar-centric frameworks, embracing a holistic approach that incorporates the full complexity of galaxies, their structure, interaction between their components, and their wider environments. The interactions between galaxies, their evolution over cosmic time, and the influence of their immediate environment play a pivotal role in shaping the regions with conditions for life development. Only by integrating these broader astrophysical factors can we form a truly comprehensive understanding of the processes that govern the emergence and sustainability of life.

\begin{acknowledgement}
The authors thank the IllustrisTNG team for making their simulations publicly available. The authors also thank the reviewer for their thoughtful and constructive feedback, which helped significantly improve the clarity, structure, and impact of a previous version of this manuscript.
\end{acknowledgement}

\paragraph{Funding Statement}

This research was supported by the Ministry of Science, Technological Development, and Innovation of the Republic of Serbia (MSTDIRS) through contract no. 451-03-136/2025-03/200002 made with the Astronomical Observatory (Belgrade).

\paragraph{Competing Interests}

None. 

\paragraph{Data Availability Statement}

The simulation data underlying this article were accessed from the publicly available IllustrisTNG simulation, available at \url{https://www.tng-project.org/data}. Although the study should be easily reproduced using publicly available data, the derived data generated in this study are available from the corresponding authors, A.M. and B.V., upon reasonable request.

\printendnotes
\printbibliography


\end{document}